# Si doping on MgB$_2$ thin films by pulsed laser deposition

Y. Zhao, M. Ionescu, J. Horvat, A. H. Li, S. X. Dou

Institute for Superconducting and Electronic Materials, University of Wollongong, NSW, 2522, Australia

Email: yz70@uow.edu.au

**Abstract:**
A series of MgB$_2$ thin films were fabricated by pulsed laser deposition (PLD), doped with various amounts of Si up to a level of 18wt%. Si was introduced into the PLD MgB$_2$ films by sequential ablation of a stoichiometric MgB$_2$ target and a Si target. The doped films were deposited at 250℃ and annealed *in situ* at 685℃ for 1min. Up to a Si doping level of ~11wt%, the superconducting transition temperature (T$_c$) of the film does not change significantly, as compared to the control, undoped film. The magnetic critical current density (J$_c$) of the film at 5K was increased by 50% for a Si doping level of ~3.5wt%, as compared to the control film. Also, the irreversibility field of Si-doped MgB$_2$ films (H$_{irr}$) at low temperature is higher than for the undoped film.

## Introduction

In order to improve the performance of the newly discovered MgB$_2$ superconductor, the enhancement of pinning force in this material is necessary. To this end, doping of MgB$_2$ in the form of bulk or tape has been carried out, using various elements or compounds [1-6]. According to the literature, Si doping gives a significant enhancement of the pinning force, while it just slightly depresses the T$_c$ value [3]. It was shown that by Si doping, Mg$_2$Si phase forms inside the MgB$_2$ matrix, and could function as a source of pinning centers [3]. It is also possible that Si may enter into the MgB$_2$ lattice as a substitute for one of the constituent atoms, causing distortion of the MgB$_2$ lattice and thus increasing the flux pinning [3].

In general, the J$_c$ of MgB$_2$ film is one order of magnitude or more higher than the J$_c$ of the bulk material. The improvement of J$_c$ in thin film was usually attributed to the much higher effective current-carrying area than in the bulk [7]. It is also presumed that the content of impurities, such as MgO, is high in the films and highly dispersed, which leads to strong pinning in the MgB$_2$ films [8,13]. The *in situ* annealed films usually have lower T$_c$, and the amount of impurities reported in these films usually appears to show an "over doped" state [9]. Another characteristic of the *in situ* annealed MgB$_2$ films is the small grain size [9, 10], which was assumed to be a consequence of the short annealing time used for their fabrication. This may lead to an increase in pinning due to the large array of grain boundaries. As was reported in the literature, the J$_c$ field dependence of the *in situ* films is weaker than for the *ex-situ* films [10], indicating a stronger pinning in the former type of film. Thus it is interesting to see if controlled Si doping could further strengthen the pinning and improve the superconducting properties in *in situ* MgB$_2$ film.

In this paper we report that a small amount Si doping in MgB$_2$ film enhances the pinning force in the *in situ* film, and improves the J$_c$ value at low temperature while gives no further depression on the T$_c$.

## Experimental details

The MgB$_2$ films were produced using a standard PLD system, in which a fixed UV laser beam (λ=248nm, 25ns) was focused on the rotating targets by cylindrical lenses. The laser energy was 300mJ/pulse, resulting in a fluence of ~3.6 J/cm$^2$ on the target, and the pulse repetition during the deposition was 10Hz. A sapphire-R (1$\bar{1}$02) substrate polished on one side, with dimensions of 5.5x2.5x0.5 mm$^3$, was attached to a resistive heater by silver paste. During deposition, the substrate temperature was kept at 250℃.

To accomplish the doping, a target-switching method was employed. In this method, three targets were mounted on a carousel. These targets were stoichiometric MgB$_2$ target (84% density), Si single crystal target, and pure Mg target. The Si doping was achieved by sequentially ablating the MgB$_2$ and the Si target 10 times during the deposition. In each round, the MgB$_2$ target was ablated for 28 sec, and the Si target was ablated for various times from 1sec to 27sec, according to the different doping levels. For the undoped sample the deposition time is 5 min. Finally, a cap layer of pure Mg was deposited onto the film surface.

The entire deposition process was carried out in high-purity Ar at 120mTorr, and only after a base vacuum of better than 1x10$^{-7}$Torr was achieved inside the deposition chamber. After the deposition process, the pressure of the high purity Ar was increased to 760Torr, and then the sample was heated to 680℃, using a heating time of 12min and maintained at that temperature for 1min. At the end of the holding time, the power was switched off, and the sample was free cooled down to room temperature at a cooling rate of about 55℃/min, following the detailed conditions given in [9].

The Si content in the film was checked by energy dispersive spectroscopy (EDS). The T$_c$ values, obtained by DC magnetization and the hysteresis loops of the films were obtained on an MPMS-5 SQUID magnetometer system (Quantum Design). The transport measurements were carried out using the standard four-probe method, on a PPMS-9T (Quantum Design) system. The applied field was perpendicular to the film surface for all the measurements. The irreversibility field ($H_{irr}$) was selected as the point where the resistivity is 10% of the value at $T_c$. The surface morphology and the thickness of the films were detected by both atomic force microscopy (AFM, Digital Instruments) and scanning electronic microscopy (SEM, Philips)
.
**Results and discussion**

The amount of Si deposited into $MgB_2$ films was calibrated on a bare substrate, using the same Si deposition conditions as during the doping of the $MgB_2$ films. The thickness of a Si film deposited at 10Hz for 300s was about 62nm, and the thickness of $MgB_2$ film deposited under the same conditions and annealed *in situ*, was about 300nm. The nominal Si content of each doped film was calculated according to this calibration.

The Si content in the $MgB_2$ films was detected by EDS, and it was calculated from the Si/Mg ratio, assuming two circumstances: a) only $MgB_2$ and Si were present; and b) only $MgB_2$ and $Mg_2Si$ phases were present in the *in situ* annealed films. The nominal Si content and the detected Si content in the $MgB_2$ films are listed in Table 1. The error in the Si content between the nominal and the detected value is large due to the probable existence of other phases, such as MgO, $MgB_4$ or excess Mg in the *in situ* annealed films. In the following text, we refer to the nominal Si content only.

For various Si doping levels, the Si deposition time varies from 1sec to 27sec. This short deposition time usually gives separated islands instead of a continuous film. This was observed by AFM on a bare substrate on which Si was deposited for 5 sec at 10Hz. As shown in Figure 1-a, the Si islands are about 100nm in diameter and several nanometers in height. Figure 1-b shows that a continuous Si film with a much longer deposition time (5 min) is built up from similar islands, which are of slightly larger diameter. As our maximum deposition time for Si was 27 sec, it is reasonable to assume that the Si introduced in the doped films is of similar discrete island structure to what is shown in Figure 1-a. In addition, in the Si doped $MgB_2$ films, a similar Si doping layer was inserted after approximately each 30nm of $MgB_2$ film, along the normal axis of the 300nm-thick film. Thus we assume that the Si is spatially homogenous inside the resulting Si doped MgB2 film.

In Figure 2 is shown a typical SEM surface aspect of the $MgB_2$ film doped with 3.5% Si. This is similar to the SEM image of the undoped film, and the roughness of the doped and undoped films obtained *in situ* is also similar.

Figure 3 shows an X-ray mapping of the Si distribution in the 3.5% and the 11% doped films. This distribution appears to be more homogeneous in the 3.5% than in the 11% Si doped film. Some Si concentrations in the 11% Si doped film represent Si droplets, and could come directly from the Si target during the deposition process. Even so, between the droplets, the distribution of Si appears to be homogenous. In the process of assessing the Si content of the films by EDS, only the surface of the films between the droplets was considered for analyses.

The zero field cooled (ZFC) and field cooled (FC) DC magnetization curves for the films with different Si content are shown in Figure 4. The $T_c$ of 1.5% Si doped $MgB_2$ film is the same as for the undoped film (27K). The $T_c$ values remain around 25K up to a doping level of 11% Si. A further increase in the Si doping level to 18% results in a decrease of $T_c$ to 23K. In contrast, in bulk $MgB_2$[3], the onset of $T_c$ is not changed for Si doping levels below 3.5%, but it appears that $T_{c\ end}$ is increased, thus making the transition narrower.

In order to reveal the comparative changes in magnetic critical current density ($J_c$) with the Si doping levels in the films, the $J_c$ was calculated from the magnetization loops, based on the critical state model. We use the standard equation [11], $J_c = 10\Delta M \times (12a/(3a-b))/b$, where $\Delta M$ ($emu/cm^3$) is the magnetization difference, a and b (cm) are the dimensions of the rectangular-shaped film (a>b). Since the MOI images of the *in situ* annealed films show homogeneous flux penetration all over our in situ film [9], this equation is valid to calculate the $J_c$ values of the films. Due to the similar dimensions of all the samples (0.55cmx0.25cmx300nm), we believe that the calculation can provide good comparison of $J_c$ between each Si doped and undoped films.

The variation of $J_c$ with the level of Si doping is shown in Figure 5. At 5K, the $J_c$ values for both the 1.5% and the 3.5% Si doped films increase relative to the undoped film. For the 3.5% Si doped film, the value of $J_c$ is the highest, approximately $10^6 A/cm^2$. This is about fifty percent higher than the values for the undoped $MgB_2$ film in fields up to 5T. In the low field regime, the $J_c$ curves show flux jump. At 10K, the improvement in $J_c$ is much smaller than at 5K, especially in high fields. The 1.5% Si doped film has the highest $J_c$, but the improvement becomes rather small. The $J_c$ is significantly decreased by 11% and 18% Si doping at all measuring temperatures (5K, 10K and 15K).

Figure 6 shows the irreversibility line and upper critical fields of the 3.5% Si doped and undoped $MgB_2$ films. Upon inspection of this figure it can be seen that the slope of the $H_{irr}$ -T curve for the 3.5% Si doped film is higher than the slope for the undoped sample, suggesting higher pinning in the former. We can also see a similar trend in the $H_{c2}$ curves of the two films.

Figure 7 shows the electric transport curves of the doped and undoped films in applied fields from 0 to 8.7 T. In the high field regime, a shift of the $T_c$ to high temperature with 3.5% Si doping is clearly seen.

However, a higher doping level of 7wt% Si draws the transition down to lower temperatures. There is a sharp increase in normal-state resistivity with further doping (370 and 700μOhm cm for the 11% and 18% Si doped films respectively). However the residual resistivity ratio (RRR) remains the same, about 1.1, for all our films. The resistivity values (130~230μOhm cm) of our low-level Si doped films are in the intermediate resistivity regime for $MgB_2$ [14].

It has been shown that among the Li, Al, Si, Zr, Ti, SiC, $Y_2O_3$, and $ZrSi_2$ doping in $MgB_2$ bulk and tapes [1-6, 12], Si is very competitive with respect to improvement in pinning force as well as in maintaining the transition temperature. In the present work, a clear improvement of $J_c$ values at low temperature in the *in situ* annealed $MgB_2$ films is also achieved by low level Si doping. Gurevich et al. have discussed the possible effects of dopants and impurities in this two-gap superconductor [13]. According to Gurevich and co-workers, the dopants substituting for Mg, such as Al and Li, add more scattering in the 3D π-band, and the dopants substituting for B, like O in our *in situ* films, can provide strong 2D σ-scattering, which may increase the slope of the $H_{c2}$-T curve. Although it is still unclear if Si would substitute for Mg or B in MgB2 lattice, the increase of the slope of the $H_{c2}$-T curve in our Si doped films indicates a further enhancement on the σ-band scattering. On the other hand, due to the reaction between Si and Mg, the Si addition in $MgB_2$ probably introduces disorder in the Mg layer as well, hence contributes to the π-band scattering, which gives an upward curvature in $H_{c2}$-T curve at lower temperature [15].

The nano-sized secondary phases introduced by Si doping, like $Mg_2Si$ embedded in the $MgB_2$ matrix [3], may improve the $J_c$ values in high fields by providing pinning centers. As discussed by Gurevich, these nano-sized particles may also produce strong out-of-plane π-scattering in $MgB_2$ [13]. The low-temperature $J_c$ improvement by Si doping in our results may to some degree result from the enhancement of π-scattering.

$MgB_2$ thin film is usually considered to have higher density and better connections between the grains compared to the bulk. For bulk doping, the increase in $J_c$ is partially attributed to the improvement in density and connection between the $MgB_2$ grains, so that a high doping level of about 10% is essential [5]. However, in the film doping circumstance, it is unlikely that the density of the film would be improved by doping. The addition of the Si provides pinning centers and improves $J_c$ only at small doping levels. With further increases of in Si content, the effective current carrying area probably decreases quickly due to the introduction of more non-superconducting phase, which leads to the sharp deterioration of $J_c$ properties in our dope $MgB_2$ films.

## Conclusions

We obtained Si-doped *in situ* annealed $MgB_2$ film by a target-switching method. The Si dopant is highly dispersed inside the film. A clear improvement in $J_c$ by Si doping in $MgB_2$ *in situ* film was achieved without obvious depression of the $T_c$. It was shown that the optimum Si doping level for $MgB_2$ thin films obtained by this method is 3.5%. The higher slope of the $H_{irr}$-T curve of Si doped film indicates an enhancement of flux pinning in the doped *in situ* annealed $MgB_2$ film.

## Acknowledgments

The authors would like to thank E. W. Collings for the kindly help of providing the stoichiometric $MgB_2$ target. This work is supported by Australian Research Council (ARC) under a Linkage Project (LP0219629) cooperating with Alphatech International and The Hyper Tech Research Inc.

Table 1. Si content in the films produced *in situ* by PLD.

| sample No. | 1 | 2 | 3 | 4 | 5 |
|---|---|---|---|---|---|
| Nominal Si content (wt%) | 1.5 | 3.5 | 5 | 11 | 18 |
| Measured Si content (wt%) -assuming $MgB_2$ +Si | 0.6 | 1.6 | 2.0 | 8.6 | 11.4 |
| Measured Si content (wt%) -assuming $MgB_2$+$Mg_2Si$ | 0.6 | 1.7 | 2.1 | 10.0 | 13.9 |

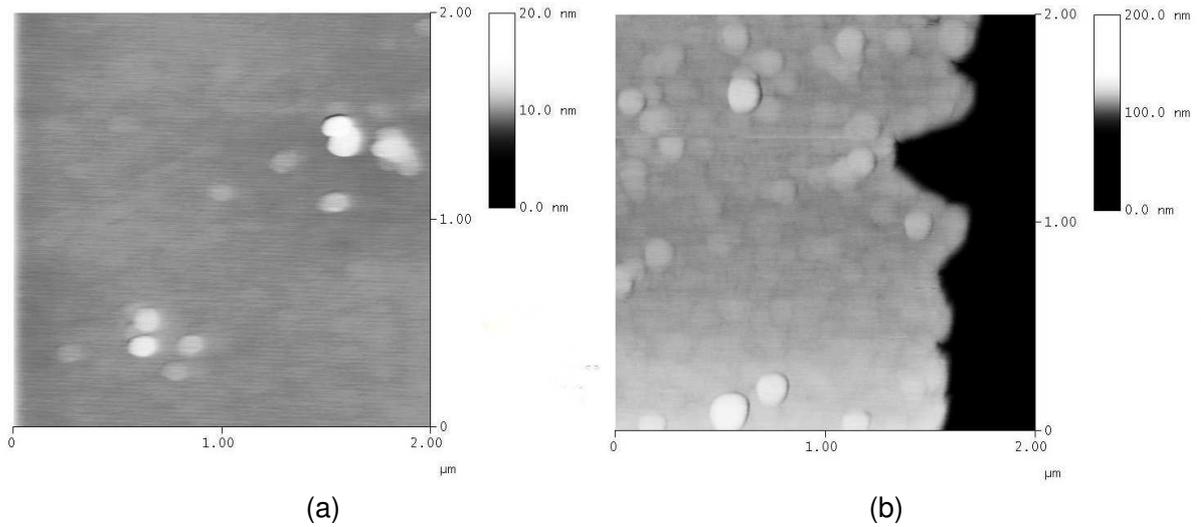

(a) (b)

Figure 1. AFM images of Si on sapphire-R substrate, a) deposited for 5sec at 10Hz, b) deposited for 5min at 10Hz. The right part of the film was intentionally scratched away, and the step revealed a film thickness of about 60 nm.

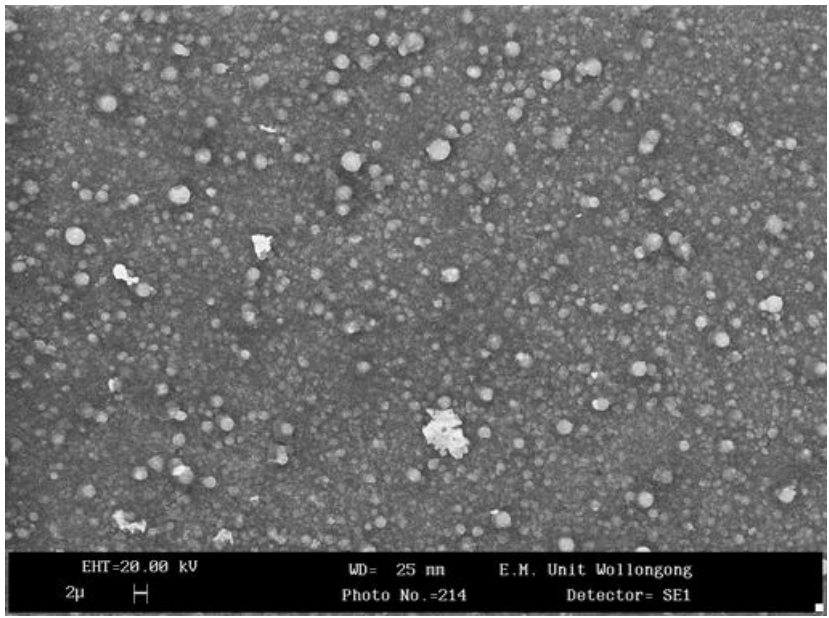

Figure 2. SEM image of 3.5%Si doped $MgB_2$ film.

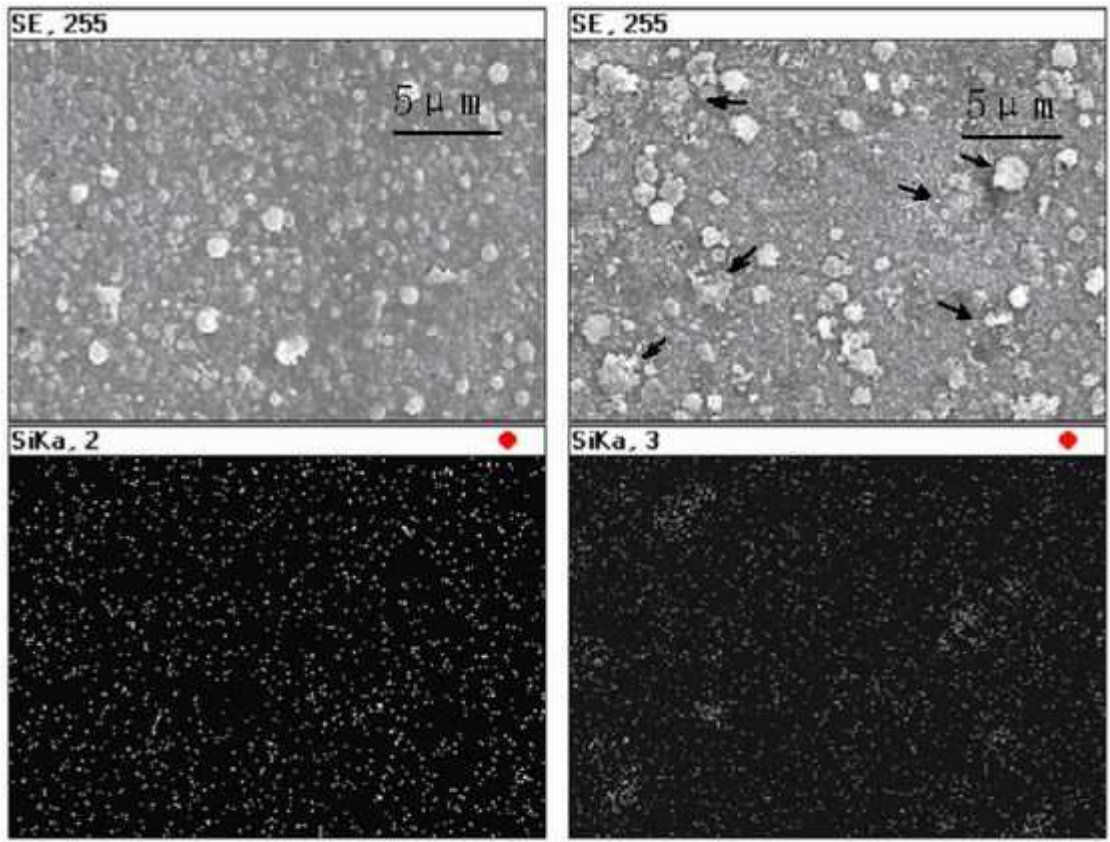

a          b

Figure 3. EDS Si mapping in the Si doped $MgB_2$ films. The upper part contains SEM secondary electron images, and the lower part the distribution of Si. a) 3.5% Si doping, b) 11% Si doping. The arrows indicate Si-rich droplets.

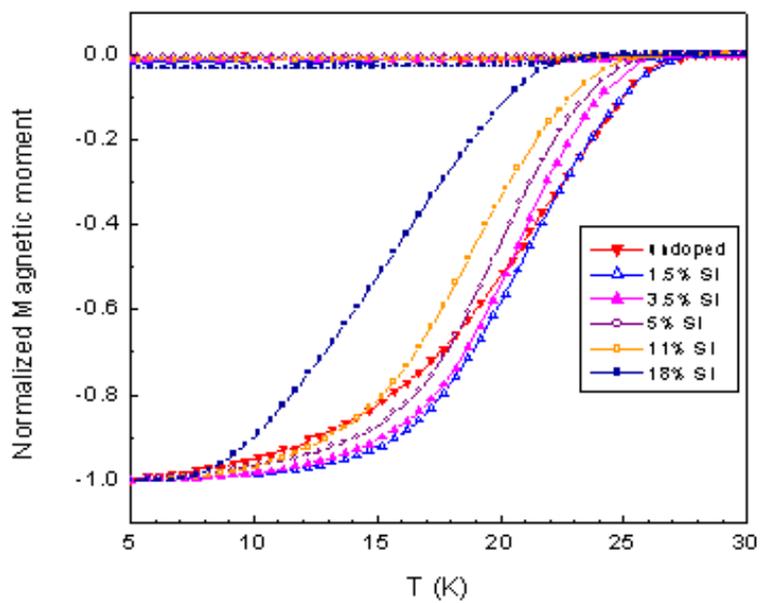

Figure 4. Magnetization curves of the films with different Si doping levels.

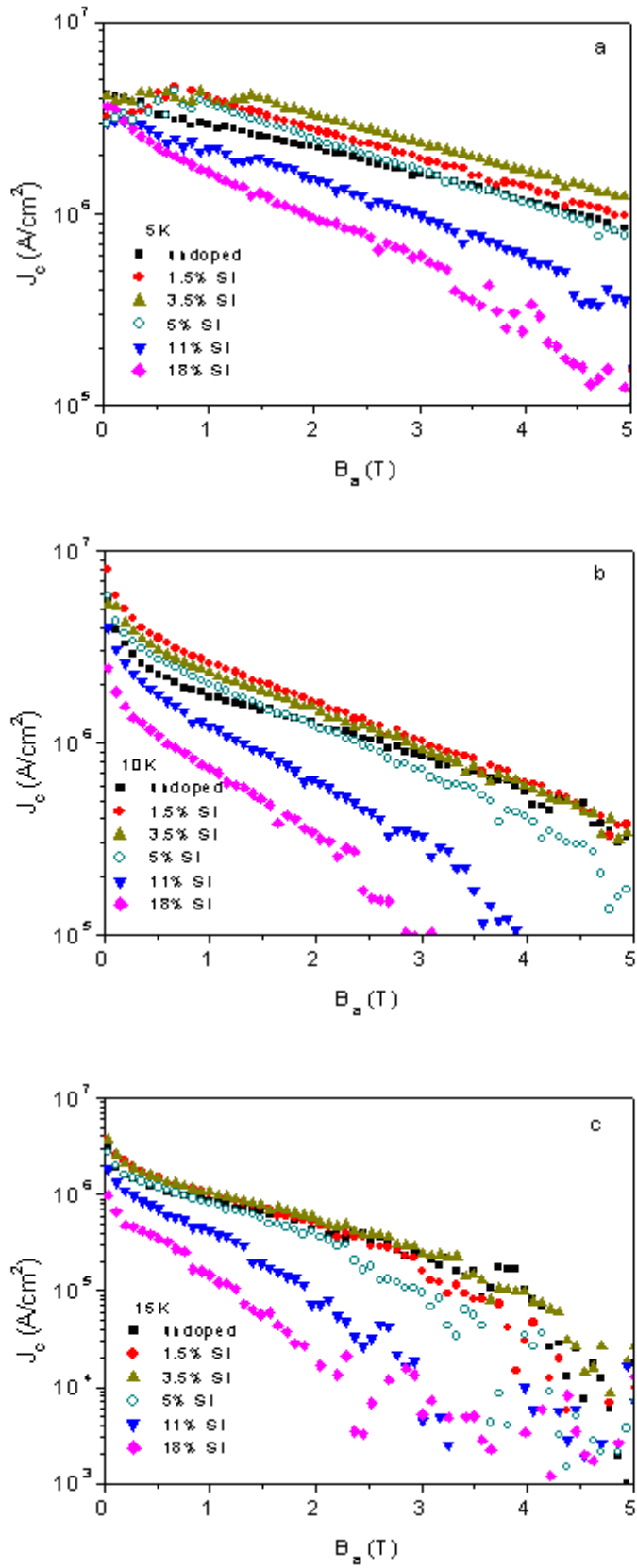

Figure 5. $J_c$ values of different Si doping levels. a: at 5K, b: 10 K, c: 15K

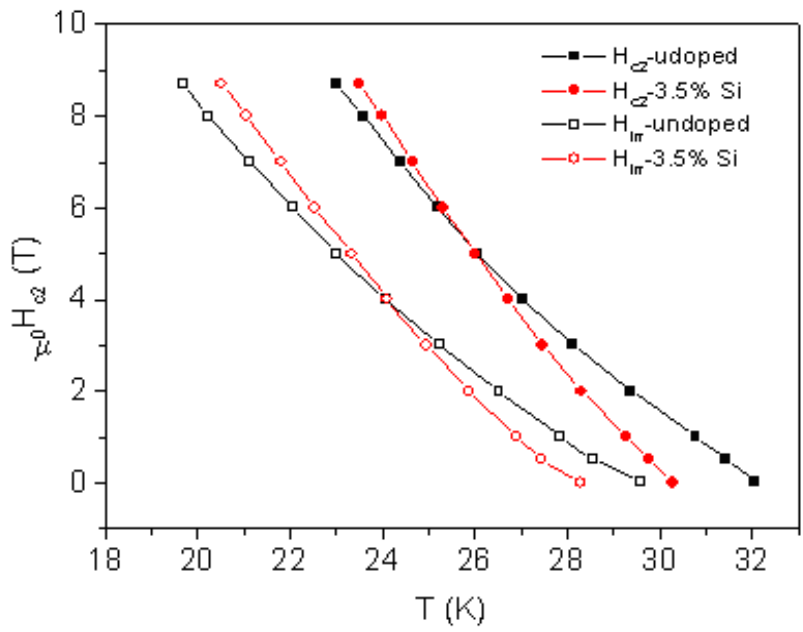

Figure 6. Irreversibility lines and upper critical fields of the 3.5% Si doped and undoped MgB$_2$ films.

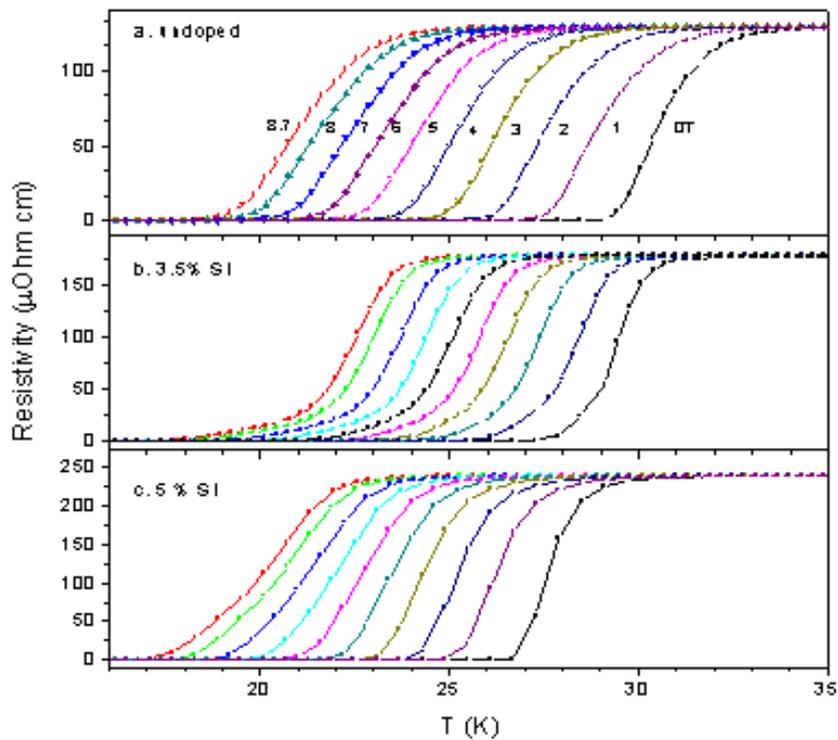

Figure 7. The resistivity versus temperature curves in fields from 0T to 8.7T. a: undoped film; b: 3.5wt% Si doped film; c: 5wt% Si doped film.